\documentclass[aps,prx,twocolumn,superscriptaddress,superscriptaddress,longbibliography]{revtex4-2}
\usepackage{graphicx} 
\usepackage[colorlinks=true, linkcolor=black, citecolor=blue, urlcolor=blue]{hyperref}
\usepackage{glossaries}
\usepackage{amsmath}
\usepackage{amsfonts}
\usepackage{mathtools}
\usepackage{bm}
\usepackage{braket}
\usepackage{graphicx}
\usepackage{makecell}
\usepackage{multirow}
\usepackage{xcolor}
\usepackage{orcidlink}

\newcommand{\quotes}[1]{``#1''}

\newacronym{qpe}{QPE}{quantum phase estimation}
\newacronym{csf}{CSF}{configuration state function}
\newacronym{sd}{SD}{Slater determinant}
\newacronym{hf}{HF}{Hartree-Fock}
\newacronym{dft}{DFT}{density functional theory}
\newacronym{ks}{KS}{Kohn-Sham}
\newacronym{scf}{SCF}{self-consistent field}
\newacronym{ao}{AO}{atomic orbital}
\newacronym{mo}{MO}{molecular orbital}
\newacronym{so}{SO}{spin orbital}
\newacronym{no}{NO}{natural orbital}
\newacronym{shci}{SHCI}{semistochastic heat bath configuration interaction}
\newacronym{sfx2c}{sf-X2C}{spin-free exact two-component}
\newacronym{ci}{CI}{configuration interaction}
\newacronym{dmrg}{DMRG}{density matrix renormalization group}
\newacronym{mps}{MPS}{matrix product state}
\newacronym{pdm}{PDM}{particle density matrix}

\begin{document}

\title{Enhancing initial state overlap through orbital optimization for faster molecular electronic ground-state energy estimation}

\newcommand{\BI}{\affiliation{
Quantum Lab, Boehringer Ingelheim, 55218 Ingelheim am Rhein, Germany}}

\newcommand{\qcware}{\affiliation{QC Ware Corporation, Palo Alto, California 94301, USA }}

\author{Pauline J. Ollitrault\orcidlink{0000-0003-1351-7546}}
\email{pauline.ollitrault@qcware.com}
\qcware

\author{Cristian L. Cortes\orcidlink{0000-0002-1163-2981}}
\qcware

\author{J\'er\^ome F. Gonthier\orcidlink{0000-0002-2933-4085}}
\qcware

\author{Robert M.~Parrish\orcidlink{0000-0002-2406-4741}}
\qcware

\author{Dario Rocca\orcidlink{0000-0003-2122-6933}}
\qcware

\author{Gian-Luca Anselmetti\orcidlink{0000-0002-8073-3567}}
\BI

\author{Matthias Degroote\orcidlink{0000-0002-8850-7708}}
\email{matthias.degroote@boehringer-ingelheim.com}
\BI

\author{Nikolaj Moll\orcidlink{0000-0001-5645-4667}}
\BI

\author{Raffaele Santagati\orcidlink{0000-0001-9645-0580}}
\BI

\author{Michael Streif\orcidlink{0000-0002-7509-4748}}
\BI

\date{\today}

\begin{abstract}

The phase estimation algorithm is crucial for computing the ground state energy of a molecular electronic Hamiltonian on a quantum computer. Its efficiency depends on the overlap between the Hamiltonian’s ground state and an initial state, which tends to decay exponentially with system size. We showcase a practical orbital optimization scheme to alleviate this issue. Applying our method to four iron-sulfur molecules, we achieve a notable enhancement, up to two orders of magnitude, compared to localized orbitals. Furthermore, our approach yields improved overlaps in cytochrome P450 enzyme models.

\end{abstract}

\maketitle

Determining the electronic ground state is at the heart of making computational predictions about molecular properties. 
While traditional classical algorithms can accurately solve this problem for many simple molecules, challenges persist for more complex systems~\cite{Santagati2024}. 
Consequently, significant research efforts are devoted to developing alternative algorithms to elucidate the ground-state properties of increasingly complex molecules. Quantum computing emerges as a particularly promising avenue in this pursuit. 
\Gls{qpe}~\cite{kitaev2002classical} stands out as the prototypical algorithm within this context, offering a pathway to efficiently determine the molecular electronic ground state.

The total cost of \gls{qpe} depends on the inverse of the weight (overlap squared), $p_0 = |\braket{\Psi_{\text{initial}}|\Psi_0}|^2$ of the initial state $\ket{\Psi_{\text{initial}}}$ with the targeted ground state $\ket{\Psi_0}$. 
This $p_0^{-1}$ scaling reflects the number of times the circuit must be repeated.
Moreover, it was demonstrated that, because of statistical errors in a single circuit run, the evolution time in each repetition also holds a $\mathcal{O}(p_0^{-1})$ dependence in the worst case, raising the total algorithm cost to $\mathcal{O}(p_0^{-2})$.~\cite{ge2019faster,lin2022heisenberg}
While recent approaches, going beyond \gls{qpe}, achieve better scalings in $p_0$~\cite{dong2022ground}, concern remains in view of the orthogonality catastrophe, i.e., the overlap of a classically tractable approximate wavefunction with the targeted exact ground state can decrease exponentially as the system size grows. 
This phenomenon was adeptly demonstrated by Lee {\em et al.}~\cite{lee2023evaluating} in their investigation of iron-sulfur cluster molecules spanning from 2 to 8 metal centers.
In the quantum computing literature, the conventional choice for the initial state is a single \gls{sd}, e.g., the \gls{hf} state. 
Lee {\em et al.}~\cite{lee2023evaluating} showed that the overlap of a single \gls{sd} with a \gls{dmrg} approximation of the ground state of the renowned FeMoco is on the order of $10^{-7}$, i.e., $10^{7}$ \gls{qpe} repetitions. State-of-the-art implementations require $\sim 10^{10}$ Toffoli gates per \gls{qpe} circuit for this system, which for optimistic error rate considerations, would take more than a day to complete~\cite{lee2021even}.  This starkly illustrates the pressing need for enhanced initial state preparation methods.

Several endeavors have been dedicated to this task, notably introducing quantum circuits for preparing a \gls{csf}, i.e., a linear combination
of \glspl{sd} defined as an eigenfunction of both the Hamiltonian and the spin operators,~\cite{carbone2022quantum, sugisaki2016quantum, sugisaki2019open} or other sums of \glspl{sd}~\cite{tubman2018postponing, ortiz2001quantum,parrish2019quantum,babbush2015chemical} instead of a single one. 
A similar idea involves the preparation of \glspl{mps}.~\cite{schon2005sequential, malz2024preparation}
Alternatively, adiabatic~\cite{aspuru2005simulated} or low-depth state preparation methods, such as the variational quantum eigensolver~\cite{peruzzo2014variational} or imaginary time evolution~\cite{Motta2020,mcardle2019variational}, are considered as possible candidates for preparing a reasonable initial state. 
They can also be augmented with boosters to effectively amplify their overlap with the ground state.~\cite{wang2022state}
Moreover, a compelling proposal for evaluating the quality of initial states via the computation of an energy distribution has recently emerged.~\cite{fomichev2023initial}

A key factor in preparing a good initial state, which has attracted less attention so far, is carefully selecting the \gls{mo} basis.
Previous studies have shown some enhancement in the overlap by working with the \glspl{no}.~\cite{tubman2018postponing,lee2023evaluating,ratini2023natural}
Interestingly, it was also demonstrated that $p_0$ converges faster than the energy, suggesting the potential for achieving accurate overlap estimates classically without requiring exhaustive knowledge of the ground state.~\cite{tubman2018postponing}
This observation prompted us to explore the feasibility of leveraging approximate ground states to refine the orbital basis, potentially surpassing the overlap achieved with \glspl{no}.
We are particularly interested in studying this effect in iron-sulfur clusters as they are found at the active sites of numerous proteins, where they play key roles in vital biological processes such as photosynthesis or nitrogen fixation. These molecules exhibit strong correlations, posing challenges for theoretical comprehension using conventional quantum chemistry approaches.~\cite{reiher2017elucidating} Hence, they serve as ideal candidates for exploration via quantum computing methodologies. Notably, they were one of the foci of the aforementioned work of Lee {\em et al.}~\cite{lee2023evaluating} on initial state preparation. To advance our understanding in this area, we aim to closely examine their findings and leverage them as foundational pillars for our research endeavors.

Even within the fault-tolerant quantum computing regime, and particularly in its early stages when qubit availability remains constrained, the mapping of fermionic systems to qubits necessitates the careful selection of a specific set of \glspl{mo} to encapsulate electronic correlation, known as the active space.~\cite{takeshita2020increasing,vorwerk2022quantum,rossmannek2023quantum} In this study, we avoid delving into the intricate process of active space selection. This topic is as important as it is complex, with ample dedicated research within the classical quantum chemistry literature.~\cite{roos1980complete,stein2019autocas,veryazov2011select}
The selection of the active space can impact both the value of $p_0$ and the resulting ground state energy, underscoring the need for meticulous consideration to uphold the chemistry of the target molecule. 
Here we leverage the extensive research history surrounding iron-sulfur clusters to select the active space.~\cite{li2019electronic, li2017spin} Our focus remains on performing orbital rotations solely within this designated space, maintaining a fixed core energy throughout our computations.\\


\emph{Orbital optimization.}
Let us consider an approximation, $\ket{\tilde{\Psi}_0} = \sum_{I}^{N_{\text{SD}}} c_{I} \ket{I}$, of the targeted ground state,  $\ket{\Psi_0}$ with real $c_I$ amplitudes. $\ket{\tilde{\Psi}_0}$ is given to us by a classically tractable procedure that we will discuss shortly. 
Our goal today is to produce a single \gls{sd}, $\ket{J'}$, in a rotated orbital basis that can be used as an efficient initial state in \gls{qpe}. 
Numerically, our task is to maximize the overlap between $\ket{J'}$ and $\ket{\tilde{\Psi}_0}$ by optimizing the \glspl{mo}, $\{\ket{\psi'_{j}} : 0\leq j \leq 2L\}$, in which $\ket{J'}$ is expressed. 
Here and throughout we use a spin-restricted formalism where the spatial parts of the $\alpha$ and $\beta$ orbitals are identical. $2L$ is the total number of orbitals in the active space since we include $L$ $\alpha$-spin \glspl{mo} and $L$ $\beta$-spin \glspl{mo}. 
The \glspl{sd} $\ket{I}$ in the expansion of $\ket{\tilde{\Psi}_0}$ are expressed in the original computational \gls{mo} basis $\{\ket{\psi_i}: 0\leq i \leq 2L\}$. 
As detailed in appendix~\ref{sec:orb_opt_app}, our task can be realized by optimizing the entries of a $L\times L$ anti-Hermitian matrix, $\hat{\kappa}$, in order to maximize 
\begin{equation}
    p_0 \equiv |\braket{J'|\tilde{\Psi}_0}|^2 = \Big| \sum_{I}^{N_{\text{SD}}} c_{I} \det\big(\hat{\mathrm{M}}^{\alpha}_{J'I}\big) \det\big(\hat{\mathrm{M}}^{\beta}_{J'I}\big) \Big|^2
    \label{eq:ovlp_det}
\end{equation}
where $\hat{\mathrm{M}}=e^{-\hat{\kappa}}$ and $\hat{\mathrm{M}}^{\sigma}_{J'I}$ is a sub-matrix of $\hat{\mathrm{M}}$ obtained by taking its rows and columns of indices $J'_{\sigma}$ and $I_{\sigma}$, respectively. $J'_{\sigma}$ and $I_{\sigma}$ are sets of indices corresponding to occupied orbitals of spin $\sigma$ in \gls{sd} $\ket{J'}$ and $\ket{I}$, respectively. \\

\emph{Overlap accuracy.}
The $p_0$ values we report in this work contain two sources of error compared to the exact weight that rules the cost of \gls{qpe}. 
The first error comes from the fact that we only have access to an approximate ground state $\ket{\tilde{\Psi}_0}$ and not the exact one, $\ket{\Psi_0}$.
Indeed, we recall that the ultimate goal is to run \gls{qpe} to find the energy of $\ket{\Psi_0}$ and that our only previous knowledge is a classically tractable approximation of it. 
The second source of error comes from truncating the sum in Eq.~\ref{eq:ovlp_det} to make the computation tractable. 

Accurate classical quantum chemistry methods rely on a compressed representation of the wavefunction to avoid storing the full amplitude statevector. One such example is the use of \glspl{mps} as in, e.g., \gls{dmrg}. In this case, the wavefunction reads
\begin{equation}
    \ket{\Psi_{\text{MPS}}} = \sum_{n_1...n_{2L}} \sum_{a_1...a_{2L-1}}^D M^{n_1}_{1a_1}M^{n_2}_{a_1a_2}...M^{n_{2L}}_{a_{2L-1}1}\ket{n_1...n_{2L}} \, .
\end{equation}
Here, the statevector is expressed as a product of $2L$ rank 3 tensors with index $n_i$ labeling the possible occupations of orbital $i$ and the other indices running up to $D$, the so-called \textit{bond dimension}. This bond dimension determines the accuracy of the representation and the cost of the algorithm. 
Another way of compressing the information is to work with a basis of \glspl{csf} rather than \glspl{sd}. A \gls{csf} is a linear combination of \glspl{sd} that allows to produce simultaneous eigenfunctions of both the Hamiltonian and the spin operator $\hat{\text{S}}^2$ (whereas the \glspl{sd} are not necessarily eigenfunctions of $\hat{\text{S}}^2$).
On the other hand, Eq.~\ref{eq:ovlp_det} implies a decompression of the information since $\det (\hat{\mathrm{M}}^{\sigma}_{J'I})$ has to be calculated for all terms in the sum. Therefore, to make the computation efficient, it is desirable to truncate the sum in Eq.~\ref{eq:ovlp_det} at the cost of introducing an error in the overlap value. 

To formally understand the significance of the above errors, let us define two general wavefunctions $\ket{\Psi}$ and $\ket{\tilde{\Psi}}$, the residual $\ket{r} = \ket{\Psi} - \ket{\tilde{\Psi}}$
and its norm $\epsilon_r = ||\ket{r}||_2 = \sqrt{\braket{r|r}}$.
Given a single \gls{sd}, $\ket{J'}$, with overlaps $\eta = \braket{J'|\Psi}$
and $\tilde{\eta} = \braket{J'|\tilde{\Psi}}$
we want to find the relation between $\eta$, $\tilde{\eta}$ and $\epsilon_r$. 
Note that we have, $\eta - \tilde{\eta} = \braket{J'|r}$.
Since $|\braket{J'|r}| \leq ||\ket{J'}||_2 \,\, ||\ket{r}||_2$, we find
\begin{equation}
    |\eta - \tilde{\eta}| \leq \epsilon_r \, .
    \label{eq:er_bound}
\end{equation}

The bound on the first source of error, coming from the fact that we only know an approximation of the true ground state, becomes 
\begin{equation}
    \epsilon_{r,1} = \sqrt{2(1- \braket{\tilde{\Psi}_0|\Psi_0})} \, .
    \label{eq:er_1}
\end{equation}
For the second type of error we note that truncating Eq.~\ref{eq:ovlp_det} is equivalent to replacing the normalized statevector $\ket{\tilde{\Psi}_0}$ by a truncated version of it $\ket{\tilde{\Psi}_{0,\text{tr}}}$, i.e., keeping only amplitudes above a certain threshold. This means that $\braket{\tilde{\Psi}_{0,\text{tr}}|\tilde{\Psi}_0}=\braket{\tilde{\Psi}_{0,\text{tr}}|\tilde{\Psi}_{0,\text{tr}}}<1$. Therefore \begin{equation}
    \epsilon_{r,2} = \sqrt{1 - \braket{\tilde{\Psi}_{0,\text{tr}}|\tilde{\Psi}_{0,\text{tr}}}}.
    \label{eq:er_2}
\end{equation} 

Clearly, for the type of systems we are targeting with \gls{qpe}, these bounds are not tight enough since $\braket{\tilde{\Psi}_0|\Psi_0}$ and $\braket{\tilde{\Psi}_{0,\text{tr}}|\tilde{\Psi}_{0,\text{tr}}}$ would tend to zero. 
However, if the orbital bases of $\ket{J'}$ and the reference state are similar, then $e^{-\hat{\kappa}}$ is close to the identity matrix. 
In other words, $\ket{J'}$ overlaps with a limited part of the reference state, i.e. only a few terms in Eq.~\ref{eq:ovlp_det} survive.
This means that we can rewrite $\eta = \braket{J'|\Psi_{\text{tr}}}$ and $\tilde{\eta} = \braket{J'|\tilde{\Psi}_{\text{tr}}}$ where $\ket{\Psi_{\text{tr}}}$ ($\ket{\tilde{\Psi}_{\text{tr}}}$) is a truncated $\ket{\Psi}$ ($\ket{\tilde{\Psi}}$) with norm much below 1. 
This directly reduces the bound to $\epsilon_r = \sqrt{\braket{\Psi_{\text{tr}}|\Psi_{\text{tr}}} + \braket{\tilde{\Psi}_{\text{tr}}|\tilde{\Psi}_{\text{tr}}} - 2 \braket{\Psi_{\text{tr}}|\tilde{\Psi}_{\text{tr}}}}$. In the next paragraphs we demonstrate this bound in several numerical examples before showing the performance of orbital optimization for iron-sulfur clusters.\\


\emph{Numerical results.}
To calculate the overlaps of interest and optimize the orbital basis, we need to find a reasonable approximation of the ground state.
In this regard, we employ either spin-adapted \gls{dmrg} from Block 2~\cite{zhai2023block2,sharma2012spin} or \gls{shci} from Dice~\cite{Holmes2016Heat,sharma2017semistochastic}. 
While the wavefunction resulting from \gls{shci} can be directly used to calculate the overlap according to Eq.~\ref{eq:ovlp_det}, the \gls{mps} ensued by \gls{dmrg}, expressed in \gls{csf} basis, needs to be post-processed. 
In this case, we employ a sampling routine to first save all \glspl{csf} with absolute amplitude above $t_{\text{CSF}}$. 
Then, we transform them to a superposition of \glspl{sd} following the algorithm of Ref.~\cite{fales2020fast}, as detailed in Appendix~\ref{sec:spin_adapt}, with a threshold $t_{\text{SD}}$ under which the \glspl{sd} are discarded.
We denote the resulting statevector as $\ket{\tilde{\Psi}_0}$. 
To obtain a new orbital basis efficiently, we first sort $\ket{\tilde{\Psi}_0}$ by order of absolute amplitude, from the largest to the smallest. 
We choose $\ket{J'}$ as the \gls{sd} with the largest amplitude in $\ket{\tilde{\Psi}_0}$. 
We optimize the orbital basis of $\ket{J'}$ using the \texttt{GradientDescent} optimizer of JAXopt~\cite{jaxopt_implicit_diff} to minimize $f(\hat{\kappa}) = 1 - |\braket{J'|\tilde{\Psi}_0}|^2$
according to Eq.~\ref{eq:ovlp_det}, with a given value $N_{\text{SD}}^{\text{opt}}$ to truncate the sum. 
We always start the optimization by setting all $\hat{\kappa}$ entries to zero, i.e., we start by expressing $\ket{J'}$ in the same basis as $\ket{\tilde{\Psi}_0}$.
All orbital rotations are strictly applied inside the active space. The core energy always remains unchanged. \\

\begin{figure}
    \centering
    \includegraphics[width=0.9\columnwidth]{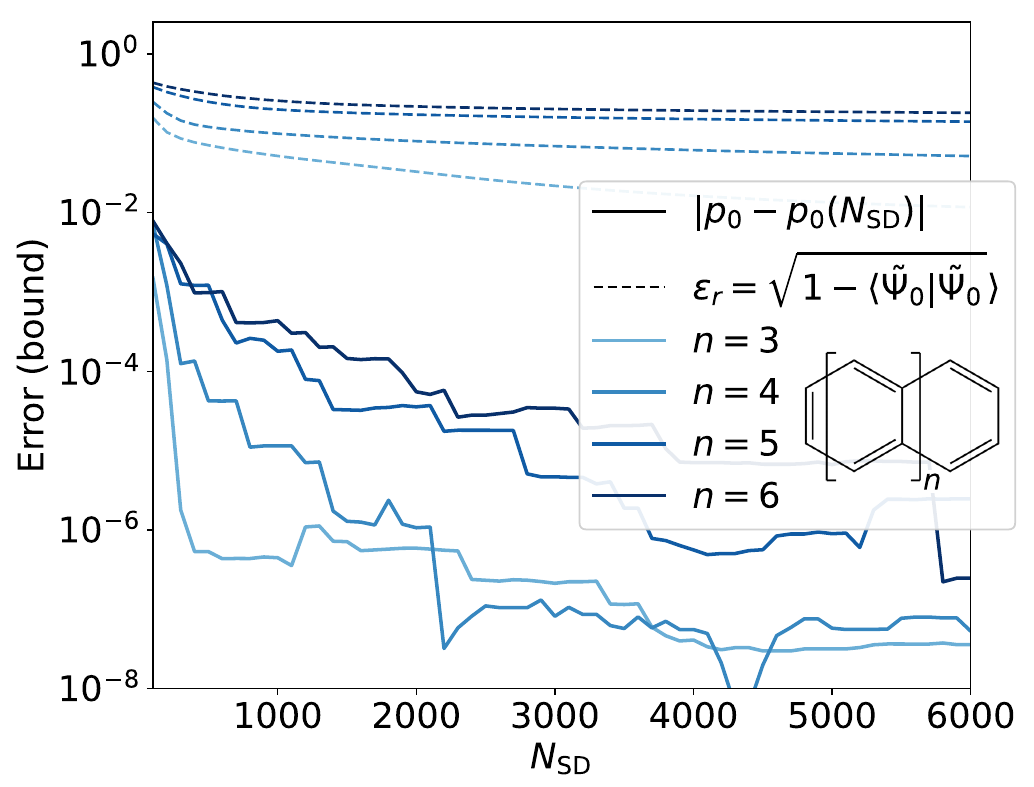}
    \caption{Convergence of $p_0$ with the number of \glspl{sd} included in the reference statevector, $\ket{\tilde{\Psi}_0}$. The full lines correspond to the error between $p_0$ calculated using $N_{\text{SD}}$ with respect to the value of the overlap obtained from all sampled \glspl{sd}. The error bound, $\epsilon_r$, is also shown in dashed lines. The results are given for acenes of different sizes $n$ and for singlet states.}
    \label{fig:acene_main}
\end{figure}

\emph{Acenes.}
As an initial exploration, we study the family of acenes, which are polycyclic aromatic hydrocarbons with a linear arrangement of fused benzene rings (see Figure~\ref{fig:acene_main}). Acenes have unique electronic properties, including high electron mobility and excellent charge transport characteristics.
As they exhibit semiconducting behavior, they are valuable in developing organic semiconductors used in various electronic devices, such as organic field-effect transistors, organic light-emitting diodes, and organic photovoltaics.
They are strongly-correlated systems and, therefore, serve as interesting models for our study. 

Following the work of Sharma {\em et al.}~\cite{sharma2019density} (using their geometries, see Appendix~\ref{sec:acenes_appendix}) we obtain the \glspl{mo} with B3LYP/6-31G(d,p). The active space corresponds to the $\pi$ electrons distributed in all valence $\pi$ orbitals leading to $2(2n+1)$ electrons in $2(2n+1)$ orbitals, where $n$ is the number of rings. 

A preliminary study on naphtalene ($n=2$), detailed in Appendix~\ref{sec:acenes_appendix}, illustrates our previous discussion on the first type of error, i.e., it indicates that the true $p_0$ error is substantially lower than $\epsilon_{r,1}$. In particular, we obtain two approximations of the singlet ground state, $\ket{\tilde{\Psi}_0^{\text{MO}}}$ and $\ket{\tilde{\Psi}_0^{\text{NO}}}$, in the \gls{mo} and \gls{no} basis, respectively. We show that, given the proximity between the two orbital bases, $|\braket{J'|\tilde{\Psi}_0^{\text{MO}}} - \braket{J'|\tilde{\Psi}_0^{\text{NO}}}| \ll ||\ket{\tilde{\Psi}_0^{\text{MO}}} - \ket{\tilde{\Psi}_0^{\text{NO}}}||_2$. This is repeated for ground states approximated with both spin-adapted \gls{dmrg} and \gls{shci}, and for several single \gls{sd} $\ket{J'}$, leading, in all cases, to the same observation. 

We also approximate the ground states of larger acenes with $3\leq n\leq 6$, in both singlet and triplet states, with spin-adapted \gls{dmrg}. We obtain, in each case, an optimized orbital basis and show that we can always increase $p_0$. In all these cases the resulting enhancement is, however, very modest. Nonetheless, this allows us to make several interesting observations. 
First, we obtain the same basis by running the optimization using 10\% or 50\% of the total number of \glspl{sd} in the reference statevector as a value for $N^{\text{opt}}_{\text{SD}}$ to truncate the sum in Eq.~\ref{eq:ovlp_det}. 
We also show that for all systems studied here, once the orbital basis is optimized and fixed, the resulting value of $p_0$ converges quickly with $N_{\text{SD}}$. 
This is highlighted in Figure~\ref{fig:acene_main} for the singlet states. We see that, for each system size, the error made in the calculated $p_0$ is substantially lower than the residual norm (corresponding to Eq.~\ref{eq:er_2}).
We also show that the optimized orbital basis differs from the \glspl{no} in all cases. The optimized basis is closer to the starting \gls{mo} basis compared to the \glspl{no}. Further detail is provided in Appendix~\ref{sec:acenes_appendix}.\\

\emph{Iron-sulfur clusters.}
As mentioned earlier, one of our focuses is on determining whether optimizing the orbital basis can significantly enhance the low overlaps observed in Ref.~\cite{lee2023evaluating}. 
Here, we focus on the clusters with the number of iron atoms $N_{\text{Fe}}=2$ and $N_{\text{Fe}}=4$. 
The active spaces are then constructed as in Ref.~\cite{lee2023evaluating}, from the full valence space of the Fe $3d$, S $3p$ and bonding ligand orbitals around each Fe atom.
The resulting active space sizes are shown in Figure~\ref{fig:fes_res}. Further details, such as the oxidation and spin states, are given in Appendix Table~\ref{tab:system_summary}. 
In the rest of this paragraph, we will denote these initial orbitals as \glspl{mo}. Although these orbitals deviate from the canonical \gls{hf} \glspl{mo}, we adopt this terminology to maintain simplicity in reading and refrain from introducing additional acronyms.

We approximate the ground states of the four clusters with spin-adapted \gls{dmrg} at bond dimension $D=8000$. 
We employ a singlet embedding~\cite{sharma2012spin} scheme for [Fe$_2$S$_2$]$^{-3}$. 
The resulting energies, given in Table~\ref{tab:FeS_summary}, are in excellent agreement with Ref.~\cite{lee2023evaluating}.
We use the converged state to optimize the orbitals which we refer to as the OPT basis. In all cases this allows us to increase the overlap by more than one order of magnitude, as shown in Figure~\ref{fig:fes_res}. Moreover, for all these clusters, optimizing the orbitals allows us to reach a higher $p_0$ than starting with a \gls{csf} in the original \gls{mo} basis. Further detail is given in Appendix~\ref{sec:fes_appendix}.

\begin{figure}
    \centering
    \includegraphics[width=1\columnwidth]{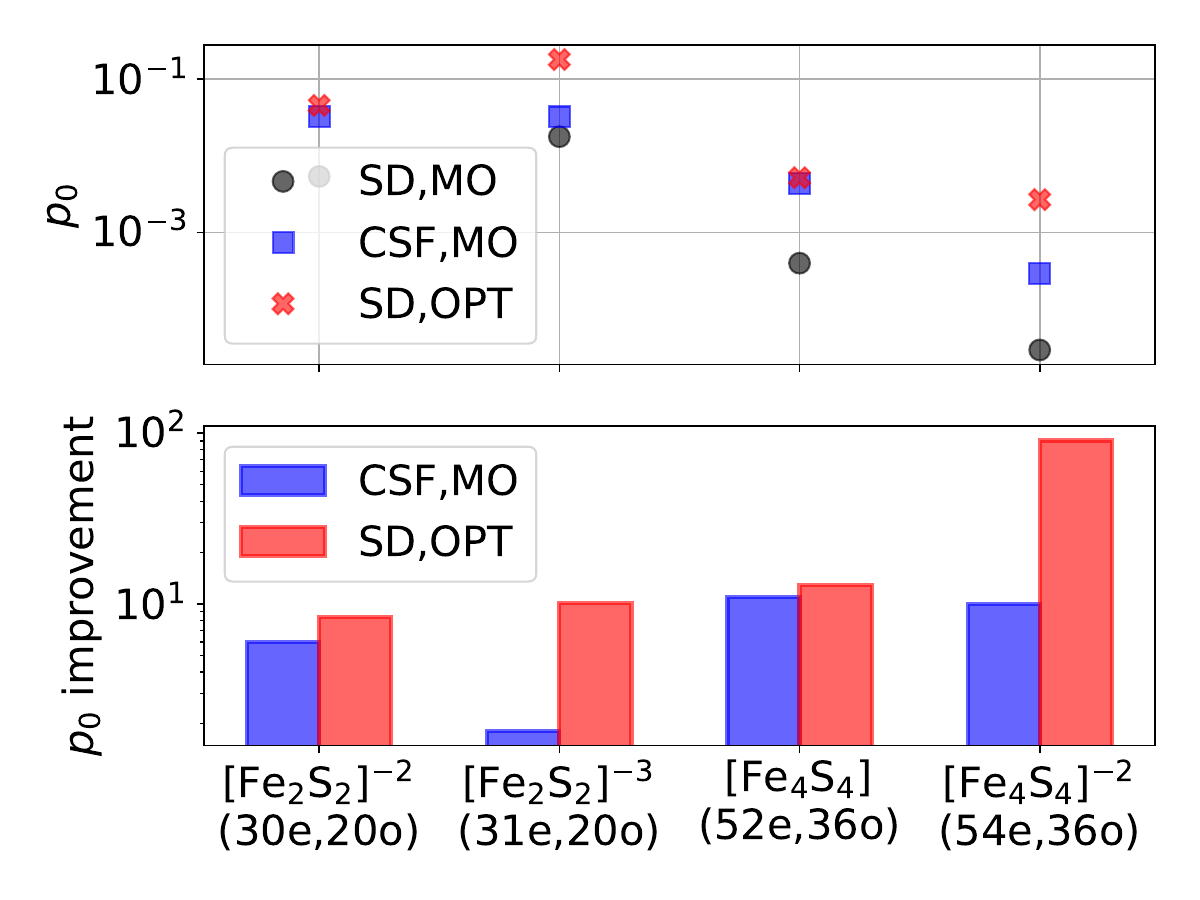}
    \caption{(Top) Weight of a single \gls{sd} or \gls{csf} expressed either in the original \gls{mo} basis or in the optimized (OPT) one for the four iron-sulfur clusters. (Bottom) Improvement in $p_0$ obtained by starting either from a \gls{csf} in the original \gls{mo} basis or from a single \gls{sd} in the optimized (OPT) basis compared to starting with a single \gls{sd} in the original \gls{mo} basis. The $x$-axis labels give the cluster name and its active space size.}
    \label{fig:fes_res}
\end{figure}

\begin{figure}
    \centering
    \includegraphics{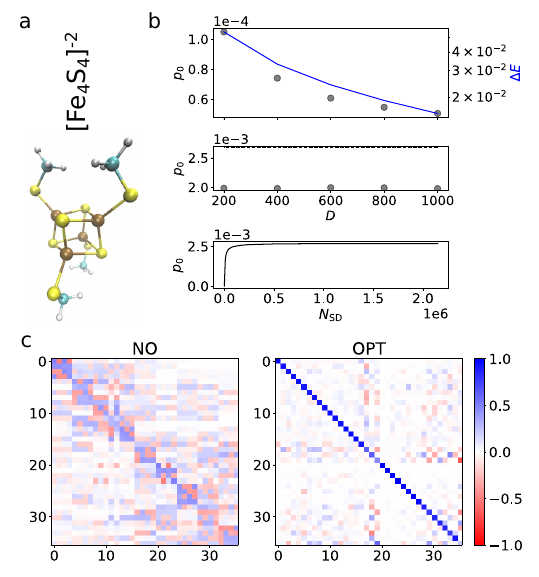}
    \caption{a. Structure of [Fe$_4$S$_4$]$^{-2}$. All the results in this Figure are obtained for this cluster. b. (Top) Convergence of the energy and weight with the \gls{dmrg} bond dimension, $D$. $\Delta E = |E(8000) - E(D)|$ is the difference between the reference energy and the one obtained at reduced $D$. Circles represent the most important absolute amplitude squared in $\ket{\tilde{\Psi}_0(D)}$. (Middle) Circles correspond to $|\braket{J'|\tilde{\Psi}_0(8000)}|^2$ where $\ket{J'}$ is in the orbital basis optimized using $\ket{\tilde{\Psi}_0(D)}$. The dashed line is $p_0$(OPT), i.e., $|\braket{J'|\tilde{\Psi}_0(8000)}|^2$ where $\ket{J'}$ is in the orbital basis optimized using $\ket{\tilde{\Psi}_0(8000)}$. (Bottom) Convergence of $p_0$(OPT) with the number of \glspl{sd} included in the reference statevector, $\ket{\tilde{\Psi}_0(8000)}$. c. Overlap matrix between the starting MOs and either the NOs or the optimized orbitals (OPT).}
    \label{fig:fe4s4_main}
\end{figure}

To show that the $p_0$ values reported in Figure~\ref{fig:fes_res} are accurate, let us focus on [Fe$_4$S$_4$]$^{-2}$. Its structure is shown in Figure~\ref{fig:fe4s4_main}a. 
First, we compare the weights obtained from different approximations of the ground state. Reference states, $\ket{\tilde{\Psi}_0(D)}$ are found with spin-adapted \gls{dmrg} at bond dimensions $D\in\{200,400,600,800,1000\}$.
We find an optimized orbital basis, OPT$D$, using each of the $\ket{\tilde{\Psi}_0(D)}$.
In Figure~\ref{fig:fe4s4_main}b (top) we show $\Delta E = |E(8000) - E(D)|$, the difference between the energies of $\ket{\tilde{\Psi}_0(8000)}$ and $\ket{\tilde{\Psi}_0(D)}$ as well as the most important amplitude squared in $\ket{\tilde{\Psi}_0(D)}$.
After the optimization, we calculate the final value of  $|\braket{J'|\tilde{\Psi}_0(8000)}|^2$ with $\ket{J'}$ in  OPT$D$, see Figure~\ref{fig:fe4s4_main}b (middle). 
Interestingly, these results clearly show that statevectors obtained at a smaller $D$, while inaccurate in energy, already allow to find an orbital basis which substantially improves $p_0$. 
In addition, we study the convergence of $|\braket{J'|\tilde{\Psi}_0(8000)}|^2$, where $\ket{J'}$ is in the OPT(8000) basis, with the truncation $N_{\text{SD}}$ of the sum in Eq.~\ref{eq:ovlp_det}. As can be seen in Figure~\ref{fig:fe4s4_main}b (bottom), $p_0$ converges very quickly with $N_{\text{SD}}$. 
In Appendix~\ref{sec:fes_appendix}, we show the same convergence behavior for all four clusters. These observations indicate that having a qualitatively correct description of the dominant part of the ground state wavefunction is enough to find an orbital basis in which $p_0$ is improved. 

Finally, we compare the \gls{no} and optimized basis. 
The NOs are obtained by diagonalizing the 1-particle density matrix of $\ket{\tilde{\Psi}_0(8000)}$. 
In Figure~\ref{fig:fe4s4_main}c, we show the overlap matrix between the starting \glspl{mo} and the \glspl{no} as well as between the starting \glspl{mo} and the optimized orbitals. As we observed in the acene case, the \glspl{no} and optimized orbitals are different, and the latter are much closer to the initial \glspl{mo}. 
Here, the direct calculation of $p_0$(NO) between the Aufbau filled \gls{sd} in the \gls{no} basis, and the \gls{mo} reference state leads to values much lower than $p_0$(MO). 
However, the density of non-diagonal elements in the \gls{mo}-\gls{no} overlap matrix is large. This translates to having more non-vanishing terms in the sum of Eq.~\ref{eq:ovlp_det}. We therefore expect the errors defined in Eqs.~\ref{eq:er_1} and~\ref{eq:er_2} to be large, hence resulting in unreliable overlap estimates. 
To allow for a more fair comparison between the OPT and \gls{no} bases, we run \gls{dmrg} again in the \gls{no} basis and get $p_0$(NO) as the amplitude squared of the most important determinant in the resulting state. 
This is only done for the two $N_{\text{Fe}}=2$ clusters, as we can afford a large enough bond dimension to compensate for the loss of locality in the orbitals. As the energy and weight converge at $D=500$, we use this bond dimension to obtain our states in the \gls{no} basis. The resulting energies are $-5092.8710$~Ha and $-5092.7187$~Ha in agreement with their counterparts in the \gls{mo} basis. 
We obtain $p_0$(NO) values of $3.35 \cdot 10^{-2}$ and $2.46 \cdot 10^{-2}$ for [Fe$_2$S$_2$]$^{-2}$ and [Fe$_2$S$_2$]$^{-3}$, respectively. In comparison the $p_0$(OPT) values are $4.53 \cdot 10^{-2}$ and $1.80 \cdot 10^{-1}$ for [Fe$_2$S$_2$]$^{-2}$ and [Fe$_2$S$_2$]$^{-3}$, respectively.
In both cases we obtain better weights in the optimized basis. \\
\emph{Cytochrome P450.}
As a final example, we apply our method to a cytochrome P450 enzyme, a system of notable relevance to the pharmaceutical industry~\cite{nelson2018cytochrome}. It has been recently demonstrated that such systems exhibit a complex electronic structure, thereby positioning them as promising candidates for exploration with future quantum computers~\cite{Goings2022}. Using our method, we successfully enhanced the initial overlap of the $S=5/2$ system described in an 43-orbital active space from $p_0=0.53$ to $p_0=0.58$. Similar results are obtained for different active space sizes and spin states. We report further details in Appendix~\ref{sec:P450}.\\
\emph{Discussion and Conclusion.}
In this work, we used orbital rotations to optimize
the overlap between a single \gls{sd} 
and 
the ground state of molecular electronic Hamiltonians. 
Specifically, we showed that for four iron-sulfur clusters, optimizing the orbital basis can lead up to $\sim$2 orders of magnitude improvement in comparison to previously predicted $p_0$ values~\cite{lee2023evaluating}.
Interestingly, expressing a single \gls{sd} in an optimized basis also improves upon the previously reported $p_0$ values between a \gls{csf} and the targeted state. Furthermore, the application of our method to the cytochrome P450 enzyme which already displays a substantially larger overlap than the iron-sulfur clusters resulted in a further 10\% increase in overlap.

Our optimization scheme leads to orbitals that resemble the starting ones, i.e., the ones in which the targeted state is approximated. This translates to a good convergence of the weight, i.e., truncating the sum in Eq.~\ref{eq:ovlp_det} to relatively low $N_{\text{SD}}$ compared to the total number of \glspl{sd} in the approximated state already leads to accurate $p_0$. 
Moreover, even states with largely unconverged energies yield improved orbital bases. This indicates that a qualitatively correct description of the dominant part of the wavefunction is sufficient to substantially enhance $p_0$.

The optimized orbitals also differ from the \glspl{no}. This is interesting as the \glspl{no} are known to minimize the orbital correlation and the entropy of the amplitudes of the statevector.~\cite{aliverti2024can}
We do not expect the optimized orbital basis to lead compact states, but this is irrelevant from the quantum computing perspective, where only the overlap between the initial and the target state matters. 

While the shown method delays the orthogonality catastrophe, it does not fully solve it. The weights calculated for the iron-sulfur clusters still demonstrate rapidly decaying behavior. Therefore, the methods discussed in this work must be combined with additional techniques to reach $p_0$ values closer to unity. 
One possibility would be to optimize the orbitals, such as to maximize the overlap of a \gls{csf} with the targeted state. 
However, classically this becomes progressively more expensive as the number of \glspl{sd} within a \gls{csf} grows exponentially with the number of its unpaired electrons.
While rotating a general quantum state into a different orbital basis is a difficult task for a classical computer, it can be done efficiently on a quantum computer by applying 
a tessellation of Givens rotation~\cite{anselmetti2021local}.
Such basis rotations were leveraged 
to reduce the runtime of \gls{qpe} by encoding factorized Hamiltonians~\cite{berry2019qubitization,von2021quantum,motta2021low,lee2021even, Goings2022, oumarou2022accelerating, rocca2024reducing}. More recently, work from Marti-Dafcik {\em et al.}~\cite{marti2024spin} introduces a compact representation of the wavefunction where different \glspl{csf} are expressed in distinct orbital bases. As highlighted in their manuscript, this ansatz is particularly well suited for preparing high overlap initial states for \gls{qpe}. 
Similarly, the orbital optimization presented here could be efficiently performed with a quantum processor. 

%

\onecolumngrid
\newpage

\appendix

\setcounter{equation}{0}

\renewcommand\theequation{A.\arabic{equation}}

\section{Orbital optimization}
\label{sec:orb_opt_app}

An orbital basis is a collection of single-particle spin \glspl{mo} defined by the product of a spatial and a spin part
\begin{equation}
    \ket{\psi_{i_{\sigma}}} = \ket{\bar{\psi}_{i_{\sigma}}} \ket{\sigma} \, .
\end{equation}
The spatial part is expressed, in turn, as a linear combination of basis functions, or \glspl{ao},
\begin{equation}
    \ket{\bar{\psi}_{i_{\sigma}}} = \sum_p^{N_{\text{AO}}} \mathcal{C}_{pi_{\sigma}}\ket{\chi_p}
\end{equation}
while the spin part is either $\ket{\alpha}$ or $\ket{\beta}$. There are then $L_{\alpha} + L_{\beta} = 2L$ \glspl{mo} in the basis.
We can write the overlap between two \glspl{mo} as
\begin{equation}
    \braket{\psi'_{j_{\tau}}|\psi_{i_{\sigma}}} = \delta_{\tau\sigma} \braket{\bar{\psi}'_{j_{\tau}}|\bar{\psi}_{i_{\sigma}}} = \delta_{\tau\sigma}\sum_{pq}^{N_\text{AO}} \mathcal{C}'_{pj_{\tau}} \mathcal{C}_{qi_{\sigma}} S_{pq}
\end{equation}
where $S_{pq} = \braket{\chi_p|\chi_q}$. 
The \gls{mo} basis, $\{\ket{\psi'_i}\}$ can be expressed as a rotation of the $\{\ket{\psi_i}\}$ basis. In other words, there exists an anti-Hermitian matrix $\hat{\kappa}$ such that 
\begin{equation}
    \mathcal{C}'_{pj_{\tau}} = \sum_{k_{\tau}}^{L_{\tau}}[e^{-\hat{\kappa}}]_{j_{\tau}k_{\tau}}\mathcal{C}_{pk_{\tau}} \, .
    \label{eq:coef_rot}
\end{equation}
Hence, the overlap between two \glspl{mo} becomes
\begin{equation}
\braket{\psi'_{j_{\tau}}|\psi_{i_{\sigma}}} = \delta_{\tau\sigma}  \sum_{pq}^{N_{\text{AO}}}\sum^{L_{\tau}}_{k_{\tau}}[e^{-\hat{\kappa}}]_{j_{\tau}k_{\tau}}\mathcal{C}_{pk_{\tau}} S_{pq} \mathcal{C}_{qi_{\sigma}} = \delta_{\tau\sigma}  \sum^{L_{\tau}}_{k_{\tau}}[e^{-\hat{\kappa}}]_{j_{\tau}k_{\tau}}\mathcal{S}_{k_{\tau}i_{\sigma}}.
    \label{eq:mo_ovlp}
\end{equation}
$\hat{\mathcal{S}}$ is the matrix containing the \glspl{mo} $\{\ket{\psi_i}\}$ mutual overlaps. Since the \glspl{mo} are orthogonal, $\hat{\mathcal{S}}$ is simply the identity matrix and hence, Eq.~\ref{eq:mo_ovlp} simplifies to
\begin{equation}
    \braket{\psi'_{j_{\tau}}|\psi_{i_{\sigma}}} =  \left\{
    \begin{array}{ll}
        [e^{-\hat{\kappa}}]_{j_{\sigma}i_{\sigma}} & \mbox{if } \tau = \sigma \\
        0 & \mbox{otherwise.}
    \end{array}
\right.
\label{eq:mo_ovlp_simple}
\end{equation}
%

A \gls{sd} is an antisymmetrized product of these orbitals. 
For a $N = N_{\alpha} + N_{\beta}$ particle system, the $N$ \glspl{mo} included in the \gls{sd} are given by two sets of indices $I_{\alpha} = \{i_{\alpha}^{\nu} : 1\leq\nu\leq N_{\alpha}\}$ and $I_{\beta} = \{i_{\beta}^{\mu} : 1\leq\mu\leq N_{\beta}\}$.
The corresponding \gls{sd} can then be defined as
\begin{equation}
    \ket{I} \equiv \mathcal{A}\big[ \prod_{\nu} \ket{\psi_{i_{\alpha}^{\nu}}} \prod_{\mu}\ket{\psi_{i_{\beta}^{\mu}}}\big]
\end{equation}
where $\mathcal{A}$ is the antisymmetrizer. 
A statevector is then defined as a linear combination of \glspl{sd}
\begin{equation}
    \ket{\Psi} \equiv \sum_{I}^{N_{\text{SD}}} c_{I} \ket{I}
\end{equation}
with real $c_{I}$  coefficients. 
The overlap between $\ket{\Psi}$ and a \gls{sd} defined in another orbital basis,
\begin{equation}
    \ket{J'} \equiv \mathcal{A}\big[ \prod_{\nu} \ket{\psi'_{j_{\alpha}^{\nu}}} \prod_{\mu}\ket{\psi'_{j_{\beta}^{\mu}}}\big],
\end{equation}
becomes
\begin{equation}
    \braket{J'|\Psi} \equiv \sum_{I}^{N_{\text{SD}}} c_{I} \braket{J'|I}
\end{equation}
The overlap between two \glspl{sd} is given by the determinant of the matrix containing all mutual orbital overlaps and thus,
\begin{equation}
\begin{split}
    \braket{J'|I} &= 
    \begin{vmatrix}
    \braket{\psi'_{j_{\alpha}^1}|\psi_{i_{\alpha}^1}} & \cdots & \braket{\psi'_{j_{\alpha}^1}|\psi_{i_{\alpha}^{N_{\alpha}}}} \\
        \vdots & \ddots & \vdots \\
        \braket{\psi'_{j_{\alpha}^{N_{\alpha}}}|\psi_{i_{\alpha}^1}} & \cdots & \braket{\psi'_{j_{\alpha}^{N_{\alpha}}}|\psi_{i_{\alpha}^{N_{\alpha}}}}
    \end{vmatrix} 
    \times 
        \begin{vmatrix}
    \braket{\psi'_{j_{\beta}^1}|\psi_{i_{\beta}^1}} & \cdots & \braket{\psi'_{j_{\beta}^1}|\psi_{i_{\beta}^{N_{\beta}}}} \\
        \vdots & \ddots & \vdots \\
        \braket{\psi'_{j_{\beta}^{N_{\beta}}}|\psi_{i_{\beta}^1}} & \cdots & \braket{\psi'_{j_{\beta}^{N_{\beta}}}|\psi_{i_{\beta}^{N_{\beta}}}}
    \end{vmatrix} \\ 
    &= \det \Big(\sum_{\nu',\nu=1}^{N_{\alpha}}\ket{\nu'}\bra{\nu} \braket{\psi_{j_{\alpha}^{\nu'}}'|\psi_{i_{\alpha}^{\nu}}}\Big) 
    \det \Big(\sum_{\mu',\mu=1}^{N_{\beta}}\ket{\mu'}\bra{\mu} \braket{\psi_{j_{\beta}^{\mu'}}'|\psi_{i_{\beta}^{\mu}}}\Big)\\
    &=\det \Big(\sum_{\nu',\nu=1}^{N_{\alpha}}\ket{\nu'}\bra{\nu} [e^{-\hat{\kappa}}]_{j_{\alpha}^{\nu'}i_{\alpha}^{\nu}}
    \Big) 
    \det \Big(\sum_{\mu',\mu=1}^{N_{\beta}}\ket{\mu'}\bra{\mu} [e^{-\hat{\kappa}}]_{j_{\beta}^{\mu'}i_{\beta}^{\mu}}) \Big)
    \end{split}
    \label{eq:ovlp_sd_detailed}
\end{equation}
where we have made use of the fact that the overlap between MOs of different spins is zero.
In order to simplify the notation, we note $\hat{\mathrm{M}}=e^{-\hat{\kappa}}$ the full $L\times L$ rotation matrix of Eq.~\ref{eq:ovlp_sd_detailed}.
The size of $\hat{\mathrm{M}}$ is $L\times L$ because we employ a spin-restricted formalism, i.e., the same matrix elements of $\hat{\mathrm{M}}$ are used for both the $\alpha$ and $\beta$ terms. 
We also introduce $\hat{\mathrm{M}}^{\sigma}_{J'I}$ the sub-matrix of $\hat{\mathrm{M}}$ obtained by taking its rows of indices $J'_{\sigma}$ and columns of indices $I_{\sigma}$.
Eq.~\ref{eq:ovlp_sd_detailed} can be re-written in matrix form as
\begin{equation}
    \braket{J'|I} = \det\big(\hat{\mathrm{M}}^{\alpha}_{J'I}\big) \det\big(\hat{\mathrm{M}}^{\beta}_{J'I}\big)
\end{equation}
and thus
\begin{equation}
    \braket{J'|\Psi} = \sum_{I}^{N_{\text{SD}}} c_{I} \det\big(\hat{\mathrm{M}}^{\alpha}_{J'I}\big) \det\big(\hat{\mathrm{M}}^{\beta}_{J'I}\big).
\end{equation}
Finally, our goal is to optimize the entries of the $\hat{\kappa}$ to minimize
\begin{equation}
    f(\hat{\kappa}) = 1 - | \braket{J'|\Psi}|^2.
\end{equation}

\section{Spin adaptation}
\label{sec:spin_adapt}

Taking advantage of spin symmetry is important for both accuracy and efficiency when the targeted molecules contain transition metals. Hence, we employ a spin-adapted version of the \gls{dmrg} algorithm.~\cite{sharma2012spin}.
After sampling, the resulting wavefunction is given as a superposition of \glspl{csf}. 
At this point we have access to the overlap of a single \gls{csf} with the ground state which is simply the maximum amplitude in the wavefunction. 
The sampled \glspl{csf} are then transformed to \glspl{sd} following the algorithm of Ref.~\cite{fales2020fast}. 
In the following we give further insight into our implementation of this algorithm. 
Let us start by saying a few words about the indexing of the \gls{sd} and \gls{csf} in the statevectors. We call $N_e$ and $N_o$ the total number of electrons and orbitals in the active space, respectively. First, the statevector is divided into seniority blocks. The seniority number, $z$, gives the number of unpaired electrons in the determinants of the corresponding block. Given a total and projected spin, $S$ and $M$, respectively, within each seniority block there are
\begin{equation}
    N_{\text{SD}}^{z} = {z \choose {z/2+M}}
\end{equation}
possible \glspl{sd} and 
\begin{equation}
    N_{\text{CSF}}^z = \frac{2S+1}{z/2 + S + 1} {z \choose {z/2 - S}}
\end{equation}
possible \glspl{csf}. We call $k_z$ the index running along the $N_{\text{SD}}^{z}$ possible \glspl{sd} or, equivalently, along the $N_{\text{CSF}}^{z}$ possible \glspl{csf}. 
Given the $k_z$ configuration, the $z$ singly occupied orbitals can be distributed in $N_o \choose z$ ways among the $N_o$ orbitals of the active space. The index along this axis is called $s_z$. Finally, given $k_z$ and $s_z$, the paired electrons are distributed in ${N_o - z} \choose {\frac{N_e - z}{2}}$ ways among the remaining orbitals. The index along this axis is $d_z$. In summary, given a seniority block $z$, a \gls{sd} or a \gls{csf} is fully determined by its indices $(k_z, s_z, d_z)$.\\
Our algorithm starts by assigning these indices to each \gls{csf} in the \gls{dmrg} output. 
For each of the existing $z, s_z, d_z$, all $N_{\text{SD}}^z$ determinants are created and their amplitude is calculated as
\begin{equation}
    A_i = \sum_{j=0}^{N^z_{\text{CSF}}} \tilde{A}_j D_{ij}
\end{equation}
where $i$ and $j$ replace the index $k_z$ used above to differentiate between \glspl{sd} and \glspl{csf}, respectively. $\tilde{A}_j$ is the amplitude of \gls{csf} $j$ and $D_{ij}$ is given by the Clebsch-Gordan coefficients (see for instance Ref.~\cite{fales2020fast} for the explicit definition of $D_{ij}$). Determinants with $|A_i|>t_{\text{SD}}$, where $t_{\text{SD}}$ is a threshold of choice, are saved in the final statevector. 

One subtlety arises when considering non-singlet systems. Since the spin-adapted \gls{dmrg} algorithm is better suited to singlet systems, a singlet embedding scheme is used to treat higher spin states~\cite{sharma2012spin}. In this case, a number (which depends on the molecule's spin) of auxiliary singly occupied orbitals need to be added to each determinant in the \gls{dmrg} output such that the overall spin becomes a singlet. The \gls{csf} to \gls{sd} transformation is then performed as explained above and the auxiliary orbitals are finally traced away in the output statevector.

\section{Acenes}
\label{sec:acenes_appendix}

In this section we study the reliability of $p_0$ estimates calculated from approximate wavefunctions. In this pursuit, we focus on acene molecules of different sizes.
The geometries of the different acenes are taken from the work of Sharma {\em et al.}~\cite{sharma2019density} and can be found in~\cite{zenodo}. 
We obtain the \glspl{mo} with B3LYP/6-31G(d,p). The active space corresponds to the $\pi$ electrons distributed in all valence $\pi$ orbitals leading to $2(2n+1)$ electrons in $2(2n+1)$ orbitals, where $n$ is the number of rings. 

\begin{figure}
    \centering
    \includegraphics[width=0.8\textwidth]{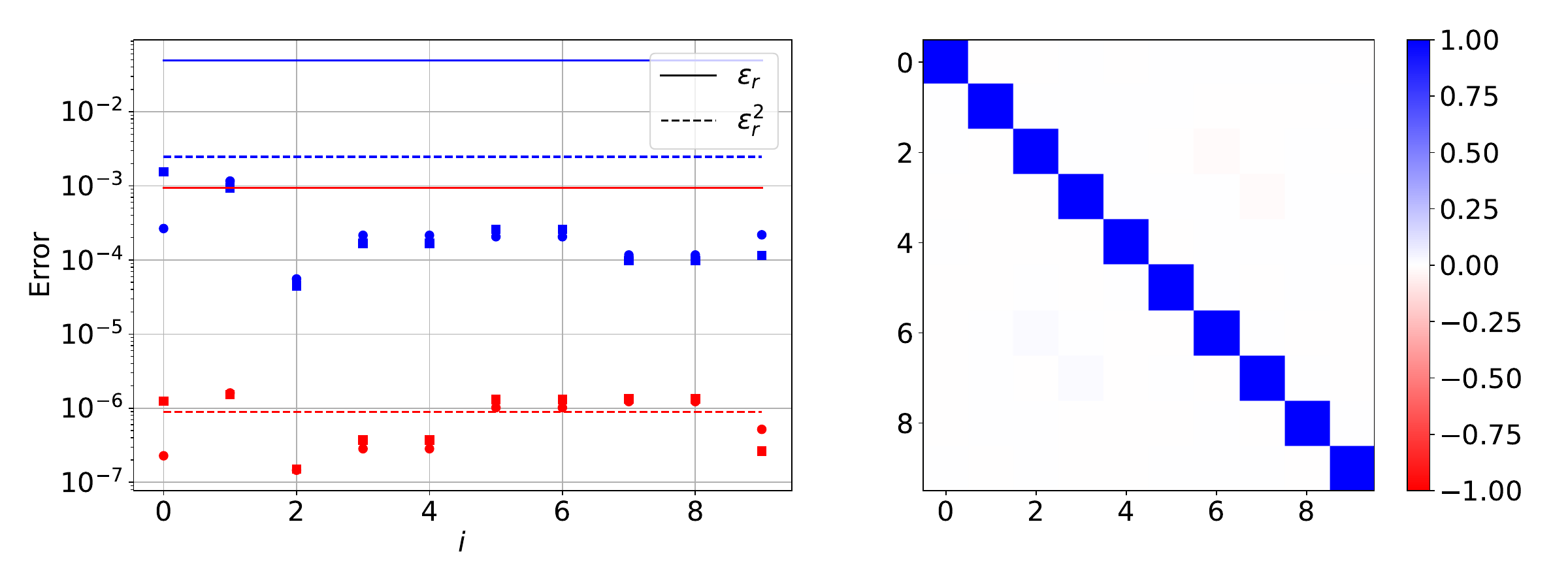}
    \caption{Results obtained with naphtalene. (left) Absolute difference between the amplitude in wavefunction A and the amplitude calculated by taking the overlap of the corresponding SD with wavefunction B. Wavefunctions A and B differ by the orbital basis in which they are expanded. Circles are for wavefunction A in \glspl{mo} and wavefunction B in \glspl{no} and squares for wavefunction A in \glspl{no} and wavefunction B in \glspl{mo}. The lines give the values of $\epsilon_r$ (and $\epsilon_r^2$), the norm of the difference between wavefunction A and B. Blue is for DMRG and red for SHCI (the two different algorithms employed to find the ground states). (right) Overlap matrix between the \glspl{mo} and \glspl{no}}.
    \label{fig:naphtalene_1}
\end{figure}

We first focus on naphtalene ($n=2$). We find an approximation of its singlet ground state with spin-adapted \gls{dmrg} at bond dimension $D=500$. We make sure that after sampling the \glspl{csf} and transformation to the \gls{sd} basis the wavefunction is normalized up to an error of $5 \cdot 10^{-7}$. 
We diagonalize the resulting 1-particle density matrix to obtain the \glspl{no}. 
The overlap matrix between the \glspl{no} and \glspl{mo} is shown on Fig.~\ref{fig:naphtalene_1} (right). We can see that they are almost identical with off-diagonal elements (2,6), (3,7), and their transpose, slightly deviating from 0. 
We run the spin-adapted \gls{dmrg} algorithm again with the same bond dimension in the \glspl{no} basis. 
In Table~\ref{tab:naphtalene_1} we show the resulting energies, number of \glspl{sd} and the weight $p_0$ of the determinant of highest amplitude, i.e., its amplitude squared. 
Despite the high similarity between the two orbital sets, $p_0$ differs in the two wavefunctions by $\sim 1.6 \cdot 10^{-3}$, indicating that a change in the orbital basis directly reflects in the value of $p_0$. 
The overlap squared between the wavefunction converged in the \glspl{mo} and \glspl{no} basis respectively is $|\braket{\tilde{\Psi}_0^{\text{MO}}|\tilde{\Psi}_0^{\text{NO}}}|^2 = 0.997521$ and indicates that \gls{dmrg} converges to slightly different states depending on the orbital basis. 
Let us write $\ket{\tilde{\Psi}_0^{\text{MO}}} = \sum_i c_i^{\text{MO}} \ket{i}$ and $\ket{\tilde{\Psi}_0^{\text{NO}}} = \sum_i c_i^{\text{NO}} \ket{i'}$ using \glspl{sd} $\ket{i}$ and $\ket{i'}$.
As per Eqs.~\ref{eq:er_bound} and~\ref{eq:er_1}, and considering that the wavefunctions are normalized, we have
\begin{equation}
    |c_i^{\text{NO}} - \braket{i'|\tilde{\Psi}_0^{\text{MO}}}| \leq \epsilon_r
\end{equation}
and
\begin{equation}
    |c_i^{\text{MO}} - \braket{i|\tilde{\Psi}_0^{\text{NO}}}| \leq \epsilon_r
\end{equation}
where
\begin{equation}
    \epsilon_r = \sqrt{2(1-\braket{\tilde{\Psi}_0^{\text{MO}}|\tilde{\Psi}_0^{\text{NO}}})}.
\end{equation}
In Figure~\ref{fig:naphtalene_1} (left), we show the errors, $|c_i^{\text{NO}} - \braket{i'|\tilde{\Psi}_0^{\text{MO}}}|$ and $|c_i^{\text{MO}} - \braket{i|\tilde{\Psi}_0^{\text{NO}}}|$, for the first 10 determinants in $\ket{\tilde{\Psi}_0^{\text{NO}}}$ and $\ket{\tilde{\Psi}_0^{\text{NO}}}$, respectively, together with $\epsilon_r$.
As anticipated in the main text, because the two orbital bases are close, this bound is very loose. Actually, the error is much closer to $\epsilon_r^2$ here. 
\begin{table}[]
    \centering
    \begin{tabular}{|c|c|c|c|c|c|}
        \hline
        Orbital basis & Method & Energy [Ha] & $N_{\text{SD}}$ & $p_0$\\
        \hline
        MOs & DMRG & -383.38958 & 1595 & 0.828714 \\
        NOs & DMRG & -383.38958 & 1578 & 0.830282 \\
        MOs & SHCI & -383.38957 & 1798 & 0.828957 \\
        NOs & SHCI & -383.38957 & 1802 & 0.830229 \\
        \hline
    \end{tabular}
    \caption{Energy, number of \gls{sd} and amplitude squared ($p_0$) of the most important \gls{sd} in the ground states converged using either the \glspl{mo} or \glspl{no} as orbital basis and \gls{dmrg} or \gls{shci} as algorithm.}
    \label{tab:naphtalene_1}
\end{table}
We repeat the same steps for ground states converged using \gls{shci} with no perturbative step and a sweep epsilon of $10^{-4}$. The results are also given in Table~\ref{tab:naphtalene_1} and Figure~\ref{fig:naphtalene_1} (in red). In this case $|\braket{\psi_0^{\text{MO}}|\psi_0^{\text{NO}}}|^2 = 0.9999991$. 
Here again, $\epsilon_r$ is a very loose bound and the errors are much closer to $\epsilon_r^2$.\\

As a second part, we shift our focus to larger acenes, namely anthracene ($n=3$), tetracene ($n=4$), pentacene ($n=5$) and hexacene ($n=6$). 
We first find the ground state in the canonical \glspl{mo} with spin-adapted \gls{dmrg}. For generality, we look for both the singlet and the triplet ground states. The convergence of the energy with the bond dimension is shown in Table~\ref{tab:energy_acenes}. The selected ground state is highlighted in bold. 
The \glspl{csf} are sampled with a threshold for sampling of $10^{-4}$. We transform them and keep only the \glspl{sd} above $10^{-4}$. The number $N_{\text{SD}}$ of \glspl{sd} in the resulting wavefunction together and its corresponding norm are shown in Table~\ref{tab:overlap_acenes}.
They form our reference states, $\ket{\tilde{\Psi}_0}$.
In the same table, we also give $p_0$(MO), the amplitude squared of the main determinant in the $\ket{\tilde{\Psi}_0}$. 

In a second step we obtain the \glspl{no} by diagonalizing the 1-particle density matrix and we calculate the weight of the Aufbau filled \gls{sd} is the \glspl{no} with $\ket{\tilde{\Psi}_0}$.
We also find a set of optimized orbitals by maximizing the weight of a single \gls{sd} with $\ket{\tilde{\Psi}_0}$. 
This single \gls{sd} is chosen to be the same as the largest amplitude one in the reference state. 
We repeat the optimization for two different truncation threshold, $N_{\text{SD}}^{\text{opt}}$, of the sum in Eq.~\ref{eq:ovlp_det}: the first one corresponds to the first 10\% \glspl{sd} in $\ket{\tilde{\Psi}_0}$ and the second one to the first 50\% \glspl{sd}. 
Once the optimized orbitals are obtained, we calculate the final value of $p_0$ between the single \gls{sd} in the optimized basis with the full $\ket{\tilde{\Psi}_0}$ (corresponding to $N_{\text{SD}}$ and norm of Table~\ref{tab:overlap_acenes}). 
This final value is reported in Table~\ref{tab:overlap_acenes}. 

\begin{figure}
    \centering
    \includegraphics[width=\textwidth]{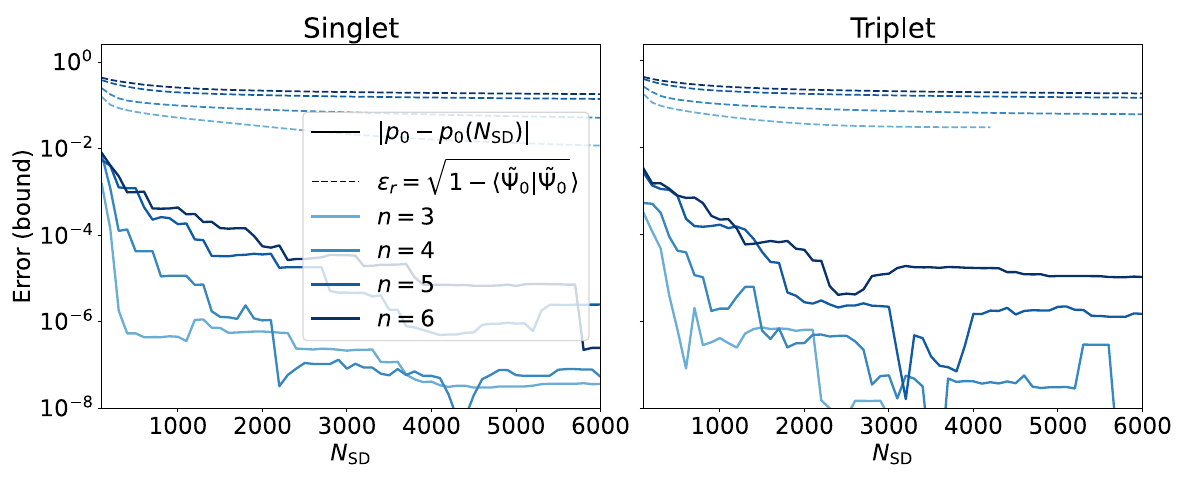}
    \caption{Convergence of $p_0$ with the number of \glspl{sd} included in the reference statevector, $\ket{\tilde{\Psi}_0}$. The full lines correspond to the error between $p_0$ calculated using $N_{\text{SD}}$ with respect to the value obtained from all sampled \glspl{sd}. The error bound, $\epsilon_r$, is also shown in dashed lines. The results are given for acenes of different sizes $n$ and for both singlet and triplet states.}
    \label{fig:acenes_ovlp}
\end{figure}

The discrepancies observed in the different overlaps are minor here. However, it is interesting to see that in all cases, the optimized orbitals lead to the largest overlap values. Moreover, and maybe more importantly, the optimizations using 10\% and 50\% of \glspl{sd} give the same result. This is an encouraging sign that $p_0$ converges fast with $N_\text{SD}$. 
To emphasize this point, we calculate $p_0$ between the \gls{sd} in the optimized basis OPT0.1 (obtained using 10\% $N_\text{SD}$) and $\ket{\tilde{\Psi}_0}$ by progressively increasing $N_{\text{SD}}$ in the sum of Eq.~\ref{eq:ovlp_det}. Figure~\ref{fig:acenes_ovlp} shows the very fast convergence of $p_0$(OPT0.1) with $N_{\text{SD}}$ for all the molecules considered here. The relative error of the weight drops below $10^{-5}$ with only 10\% of the total number of \glspl{sd} for all the systems considered here. 

The $p_0$(NO) values are unexpectedly lower than the corresponding $p_0$(MO). However, as we discussed above and numerically witnessed in the case of naphtalene, these overlaps can be tainted by a large error due to divergence in the reference states obtained in one or the other basis and due to truncating of the sum in Eq.~\ref{eq:ovlp_det}. 
In fact, if we approximate the ground state again, in the \gls{no} basis, with \gls{dmrg} (same bond dimension as for the \gls{mo} state), the resulting $p_0$(NO) values, taken as the most important amplitude squared in the statevector, are very close to the ones obtained with the optimized orbitals. 

\begin{figure}
    \centering
    \includegraphics[width=1\textwidth]{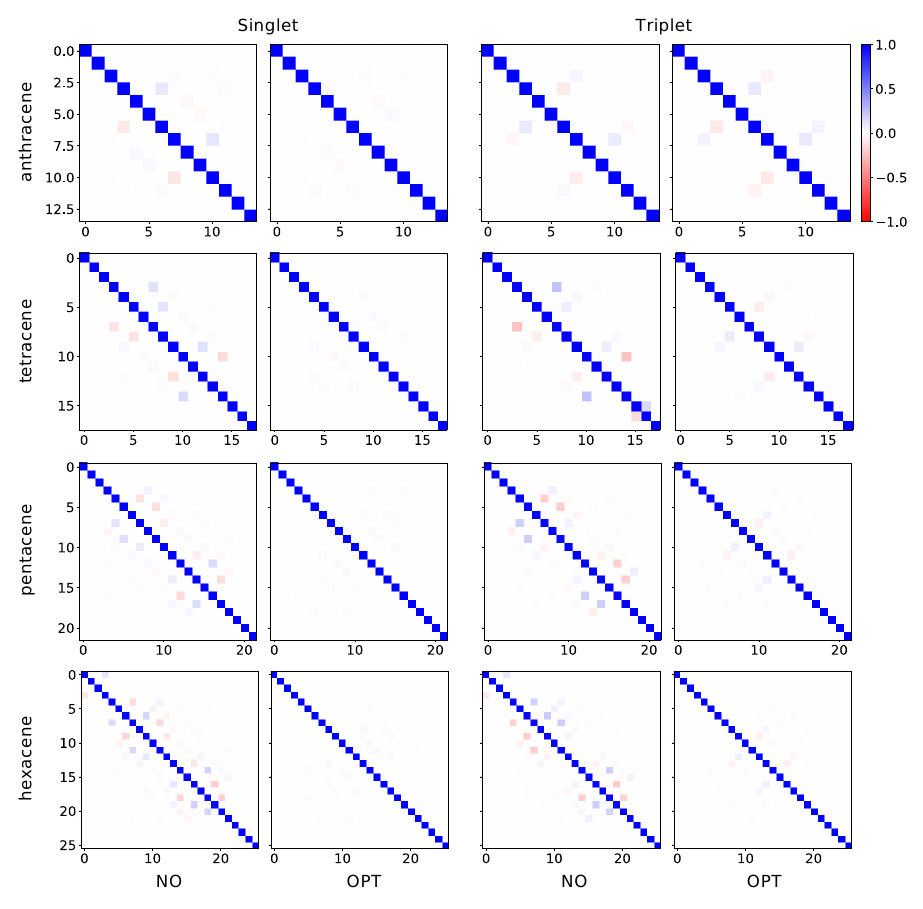}
    \caption{Overlap matrix between the starting \glspl{mo} and either the \glspl{no} or the optimized orbitals (OPT) obtained with the first 10\% of the \glspl{sd} in the reference state for each acenes in both singlet and triplet states.}
    \label{fig:acenes_s12}
\end{figure}

In Figure~\ref{fig:acenes_s12}, we compare the \glspl{no} and optimized orbitals (obtained with 10\% of the \glspl{sd}) by showing their overlap matrix with the original \glspl{mo}. We clearly see that they are distinct orbital sets, the optimized ones remaining closer to the \glspl{mo}. However, it is interesting to remark that the two types of orbitals also hold similarities, i.e., we observe a common structure in their overlap matrix with the original \glspl{mo}.

\begin{table}[]
    \centering
    \begin{tabular}{|c|c|c|c|}
        \hline
        anthracene (14,14) & tetracene (18,18) & pentacene (22,22)& hexacene (26,26) \\
        \hline
        \textbf{D=500 -536.049037} & D=500 -688.687930 & D=1000 -841.361452 & D=1500 -993.998918\\
        D=1000 -536.049038 & \textbf{D=1000 -688.687953}& D=1500 -841.361630 & D=2000 -993.999350\\
        & D=1500 -688.687954 & \textbf{D=2000 -841.361700} & \textbf{D=2500 -993.999574}\\  
        \hline
        \textbf{D=500 -535.975915} & \textbf{D=1000 -688.622454} & D=1500 -841.320648 &D=2000 -993.959774 \\
        D=1000 -535.975915 & D=1500 -688.622460& \textbf{D=2000 -841.320829} & \textbf{D=2500 -993.960074}\\ 
        \hline
    \end{tabular}
    \caption{Convergence of the energy of different acenes, with active space (number of electrons,number of orbitals), obtained with DMRG at bond dimension $D$. The upper and lower rows are for singlet and triplet states, respectively.}
    \label{tab:energy_acenes}
\end{table}

\begin{table}[]
    \centering
    \begin{tabular}{|c|c|c|c|c|c|c|c|c|}
    \hline
        Molecule & S & Energy [Ha] & $N_{\text{SD}}$ & Norm & $p_0$(MO) & $p_0$(NO) & $p_0$(OPT0.1) & $p_0$(OPT0.5)\\
        \hline
        anthracene & 0 & -536.049037 & 7522 & 0.99995 & 0.81655 & 0.81481 & 0.81892 & 0.81892 \\
        tetracene & 0 & -688.687953 & 13029 & 0.99909 & 0.76942 & 0.76833 & 0.77301 & 0.77301 \\
        pentacene & 0 & -841.361700 & 50842 & 0.99506 & 0.67247 & 0.66881 & 0.67774 & 0.67774 \\
        hexacene & 0 & -993.999574 & 64098 & 0.99207 & 0.61155 & 0.60896 & 0.61800 & 0.61800 \\
        \hline
        anthracene & 1 & -535.975916 & 4230 & 0.99959 & 0.78481 & 0.76678 & 0.79843 & 0.79843 \\
        tetracene & 1 & -688.622454 & 10208 & 0.99847 & 0.73751 & 0.72471 & 0.74703 & 0.74703 \\
        pentacene & 1 & -841.320814 & 29829 & 0.99308 & 0.67598 & 0.66598 & 0.68507 & 0.68506 \\
        hexacene & 1 & -993.960030 & 38755 & 0.98923 & 0.62550 & 0.61755 & 0.63316 & 0.63316 \\
        \hline
    \end{tabular}
    \caption{Results for the four studies acenes in both singlet and triplet states. The ground state is approximated with spin-adapted DMRG and leads the reported energies. After sampling and transforming to the \gls{sd} basis, the statevectors contain $N_{\text{SD}}$ \glspl{sd} and have the reported norm. $p_0$(MO) is the amplitude squared of the most important determinant. $p_0$(NO) is the weight of the Aufbau filled determinant in the \gls{no} basis with the statevector. $p_0$(OPT$X$) is the weight of the most important determinant in the reference statevector expressed in the optimized orbital basis obtained by using the first $XN_{\text{SD}}$ \glspl{sd}.}
    \label{tab:overlap_acenes}
\end{table}

\section{Iron-Sulfur clusters}
\label{sec:fes_appendix}

One of our focuses is on determining whether optimizing the orbital basis can significantly enhance the weak overlaps observed in Ref.~\cite{lee2023evaluating}. To maintain a seamless progression in our research, we diligently replicate the findings of Ref.~\cite{lee2023evaluating} and build upon them. 
Here we focus on the clusters with a number of iron atoms $N_{\text{Fe}}=2$ and $N_{\text{Fe}}=4$. 
Tab.~\ref{tab:system_summary} provides details about the studied iron-sulfur clusters, supplying all the essential information for our investigation.

In the following paragraph, we describe the procedure followed to obtain the active space, which is the same as the one of Ref.~\cite{lee2023evaluating}. We start by running a high spin unrestricted \gls{dft} calculation with PySCF~\cite{sun2018pyscf}. The specific spin numbers, as well as information about basis sets and functionals, can be found in Tab.~\ref{tab:system_summary}. For all systems we use the \gls{sfx2c} Hamiltonian. 
\begin{table}[ht]
    \centering
    \begin{tabular}{|c|c|c|c|c|c|c|c|}
    \hline
    \hline
    molecule & oxidation &charge & high S & low S & AS& functional & basis set\\
    \hline
    &&&&&&&\\
    \makecell{[Fe$_2$S$_2$]$^{-3}$\\ \cite{sharma2014low}}&Fe$^{\text{II}}$, Fe$^{\text{III}}$ & -3 & 9/2 & 1/2 & (31e,20o) & BP86 & TZP-DKH\\
    &&&&&&&\\
    \makecell{[Fe$_2$S$_2$]$^{-2}$\\ \cite{sharma2014low}}&Fe$^{\text{III}}$, Fe$^{\text{III}}$ &-2 &  5 & 0 &(30e,20o)& BP86 & TZP-DKH\\
    &&&&&&&\\
    \makecell{[Fe$_4$S$_4$]$^{-2}$\\ \cite{sharma2014low}}&2Fe$^{\text{II}}$, 2Fe$^{\text{III}}$ & -2 & 9 & 0 & (54e,36o)& BP86 & TZP-DKH\\
    &&&&&&&\\
    \makecell{[Fe$_4$S$_4$]}&4Fe$^{\text{III}}$ & 0 & 10 & 0 &(52e,36o) & BP86 & TZP-DKH\\ 
    &&&&&&&\\
    \hline
    \end{tabular}
    \caption{Details about the studied iron-sulfur clusters}
    \label{tab:system_summary}
\end{table}

\begin{figure}
    \centering
    \includegraphics[width = 0.9\textwidth]{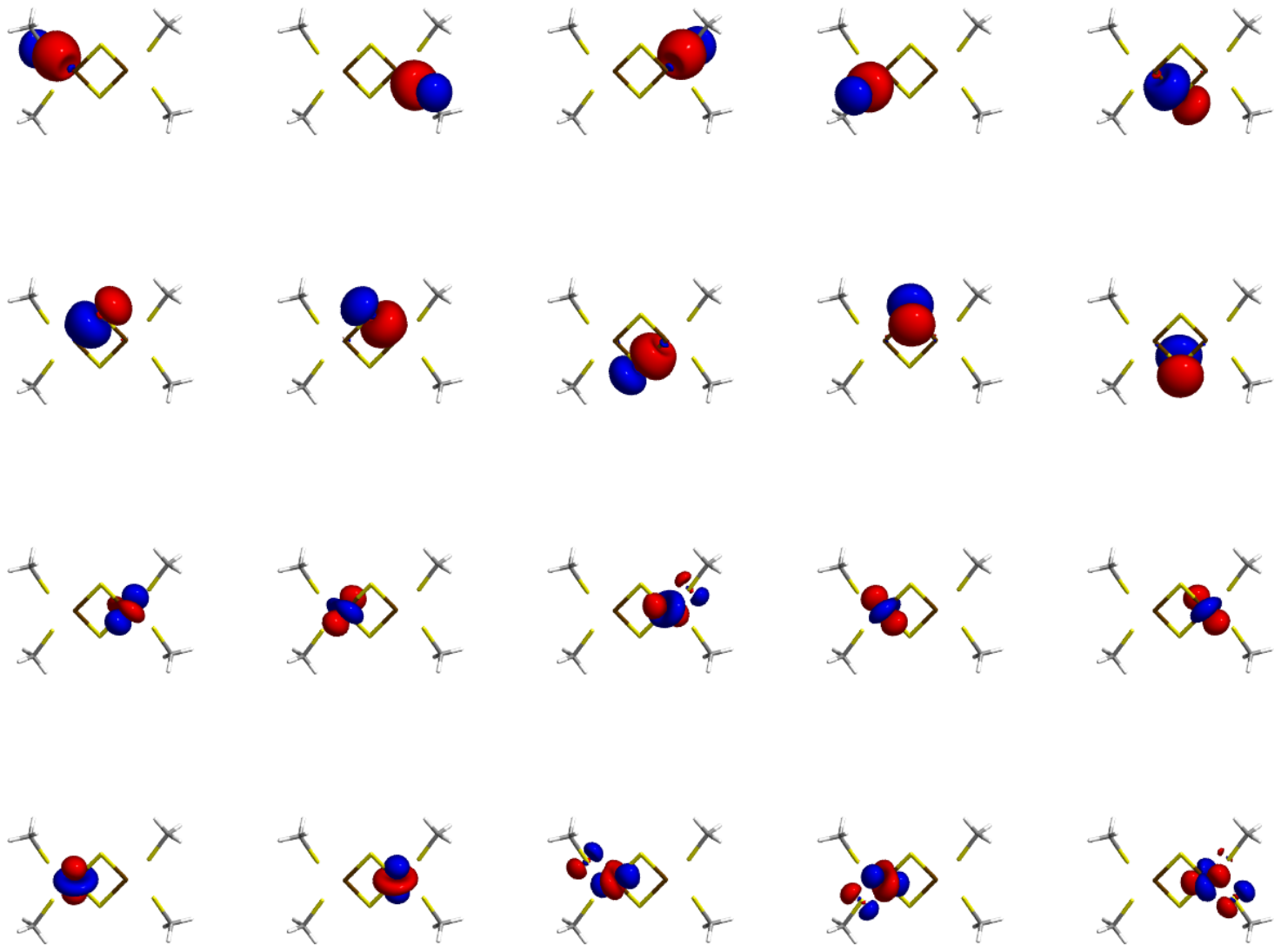}
    \caption{Pictures of the orbitals in the active space of [Fe$_2$S$_2$]$^{-2}$.}
    \label{fig:fe2s2_lmos}
\end{figure}

The spin averaged density matrix, $\rho = 1/2 (\rho_{\alpha} + \rho_{\beta})$ is calculated in the basis of the resulting \gls{ks} orbitals and transformed to the Lowdin basis, $\rho' = S^{1/2} \rho S^{1/2}$. 
The \glspl{no} and corresponding occupation numbers are obtained by diagonalizing $\rho'$. 
The \glspl{no} are transformed to the non-orthogonal \gls{ao} basis and classified in a core, active, or virtual group of orbital according to their occupation number, i.e., orbitals with occupation number above 1.95, between 0.05 and 1.95 and below 0.05 are classified as core, active and virtual respectively.
The core and active orbitals are then split localized with the Pipek-Mezey algorithm. 
The active spaces are then constructed as in Ref.~\cite{lee2023evaluating}, from the full valence space of the Fe 3d, S 3p, and bonding ligand orbitals around each Fe atom. 
These active orbitals are localized one more time and shown for [Fe$_2$S$_2$]$^{-2}$ in Figure~\ref{fig:fe2s2_lmos}. The orbital pictures for all systems are available in~\cite{zenodo}. The resulting active space sizes are given in Tab.~\ref{tab:system_summary}.

We approximate the ground states of the four clusters with spin-adapted \gls{dmrg} at bond dimension $D=8000$. 
We employ a singlet embedding~\cite{sharma2012spin} scheme for [Fe$_2$S$_2$]$^{-3}$. 
The resulting energies, given in Table~\ref{tab:FeS_summary}, are in close agreement with Ref.~\cite{lee2023evaluating}.
We sample the \glspl{csf} with $t_{\text{CSF}} = 10^{-4}$ for $N_{\text{Fe}}=2$ and $t_{\text{CSF}} = 5 \cdot 10^{-4}$ for $N_{\text{Fe}}=4$. We transform them and discard all \glspl{sd} with absolute amplitude below $10^{-4}$. The resulting number of \glspl{sd} in the statevector, $\ket{\tilde{\Psi}_0}$, and its norm are summarized in Table~\ref{tab:FeS_summary}.
In all cases, we sort these reference statevectors by decreasing order in absolute amplitude.
We optimize the orbitals using the first 10\% \glspl{sd} and the final value we report in Table~\ref{tab:FeS_summary} is $p_0(\text{OPT}) = |\braket{J'|\tilde{\Psi}_0}|^2$ where $\ket{J'}$ is expressed in the optimized basis and $\ket{\tilde{\Psi}_0}$ (expressed in the original \glspl{mo}) contains the full set of sampled \glspl{sd} corresponding to $N_{\text{SD}}$ of Table~\ref{tab:FeS_summary}. 
The orbital optimization increases the weight by more than one order of magnitude in all cases. 
Moreover, for all these clusters, optimizing the orbitals allows to reach a higher $p_0$ than starting with a \gls{csf}. As a matter of comparison, the \gls{csf} of largest amplitude in [Fe$_4$S$_4$]$^{-2}$ comprises more than 48000 \glspl{sd}. \\

\begin{table}[]
    \centering
    \begin{tabular}{|c|c|c|c|c|c|c|c|}
    \hline
        Molecule & Energy [Ha] & Norm & $N_{\text{SD}}$ & $p^{\text{CSF}}_0$(MO) & $p_0$(MO) & $p_0$(OPT) & $p_0$(NO)\\
    \hline
         \makecell{[Fe$_2$S$_2$]$^{-2}$} & -5092.8710 & 9.85e-1 & 3121044 & 3.21e-2 & 5.35e-3 & 4.53e-2 & 3.36e-2\\
         \makecell{[Fe$_2$S$_2$]$^{-3}$} & -5092.7187 & 9.94e-1 & 734408 & 3.19e-2 & 1.76e-2 & 1.80e-1 & 2.46e-2\\
         \makecell{[Fe$_4$S$_4$]$^{-2}$} & -8432.8070 & 3.65e-1 & 2146432 & 2.94e-4 & 2.94e-5 & 2.67e-3 & -\\
         \makecell{[Fe$_4$S$_4$]} & -8432.6267 & 3.65e-1 & 2127398 & 4.37e-3 & 3.97e-4 & 5.16e-3 & -\\
    \hline
    \end{tabular}
    \caption{Energy, norm characteristics and overlaps of the iron-sulfur clusters. The ground state is approximated with spin-adapted \gls{dmrg} ($D = 8000$) and leads to the reported energies. After sampling and transforming to the \gls{sd} basis, the statevectors contain $N_{\text{SD}}$ \glspl{sd} and have the reported norm. $p^{\text{CSF}}_0$(MO) and $p_0$(MO) are the amplitude squared of the most important \gls{csf} and \gls{sd}, respectively. $p_0$(OPT) is the overlap of the most important \gls{sd} in the reference statevector expressed in the optimized orbital basis obtained by using the first 10\% \glspl{sd}. $p_0$(NO) is the amplitude squared of the most important determinant in the state resulting from running \gls{dmrg} in the \gls{no} basis (see main text).}
    \label{tab:FeS_summary}
\end{table}

\begin{figure}
    \centering
    \includegraphics[width = 0.7 \textwidth]{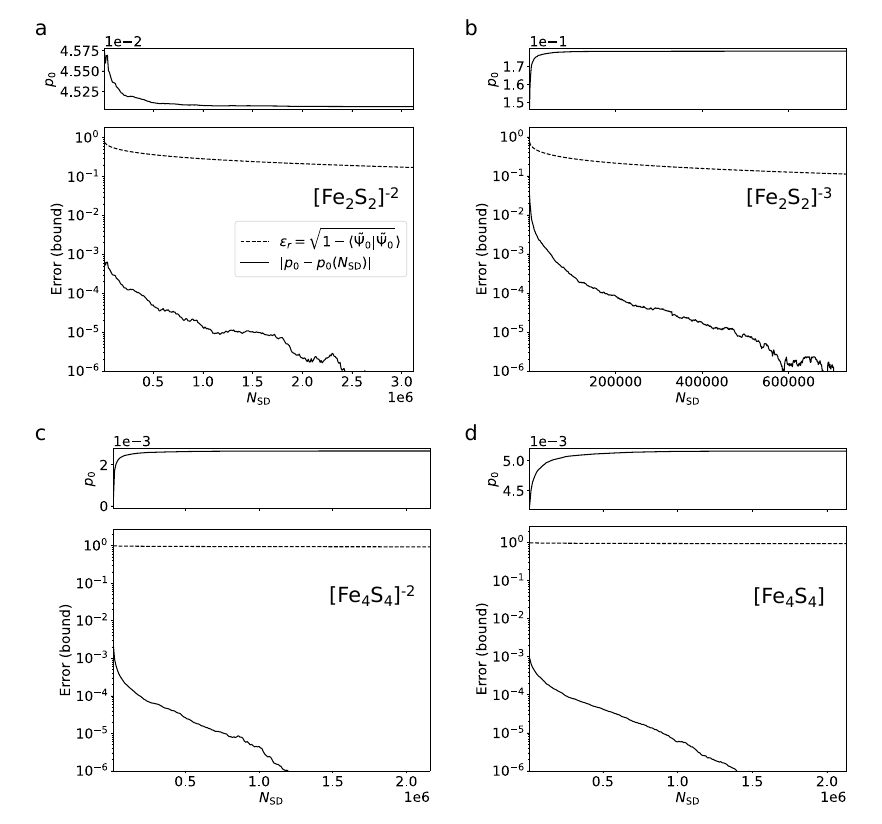}
    \caption{Convergence of $p_0$(OPT) with the number of \glspl{sd} included in the reference statevector, $\ket{\tilde{\Psi}_0}$. The full line corresponds to the error between $p_0$(OPT) calculated using $N_{\text{SD}}$ with respect to the value of the weight obtained from all sampled \glspl{sd}. The error bound $\epsilon_r$ is shown in dashed line. Plots a, b, c and d are for [Fe$_2$S$_2$]$^{-2}$, [Fe$_2$S$_2$]$^{-3}$, [Fe$_4$S$_4$]$^{-2}$ and [Fe$_4$S$_4$], respectively.}
    \label{fig:fes_Nsd}
\end{figure}

\begin{figure}
    \centering
    \includegraphics[width = 0.9 \textwidth]{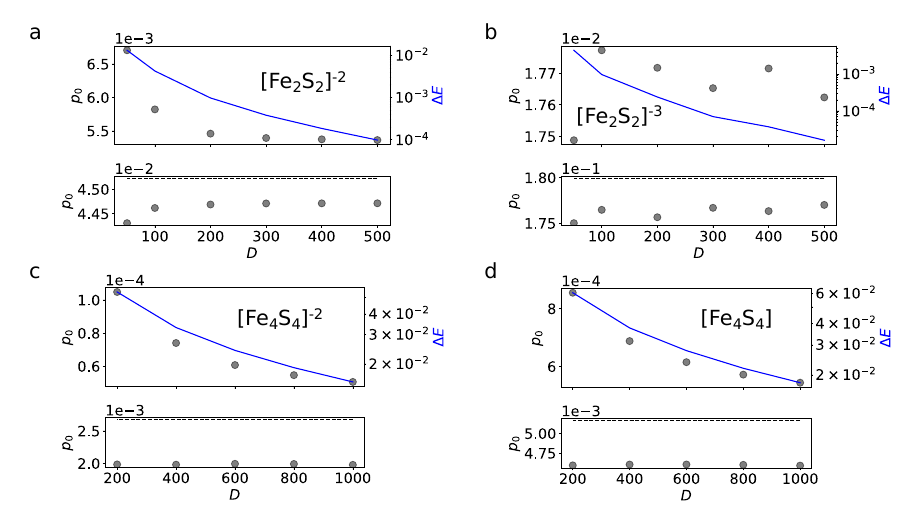}
    \caption{(Top) Convergence of the energy and weight with the \gls{dmrg} bond dimension, $D$. $\Delta E = |E(8000) - E(D)|$ is the difference between the reference energy and the one obtained at reduced bond dimension $D$. Circles represent the most important absolute amplitude squared in $\ket{\Psi(D)}$. (Bottom) Circles correspond to $|\braket{J'|\Psi(8000)}|^2$ where $\ket{J'}$ is in the orbital basis optimized using $\ket{\Psi(D)}$. The dashed line is $p_0$(OPT), i.e., $|\braket{J'|\Psi(8000)}|^2$ where $\ket{J'}$ is in the orbital basis optimized using $\ket{\Psi(8000)}$. 
    Plots a, b, c and d are for [Fe$_2$S$_2$]$^{-2}$, [Fe$_2$S$_2$]$^{-3}$, [Fe$_4$S$_4$]$^{-2}$ and [Fe$_4$S$_4$], respectively.}
    \label{fig:fes_D}
\end{figure}

As for the acenes, we study the convergence of the weight with $N_{\text{SD}}$.
The results are shown in Figure~\ref{fig:fes_Nsd}. We see again that $p_0$(OPT) converges very quickly with $N_{\text{SD}}$. 
We also study the effect of using an approximate wavefunction to optimize the orbitals. 
To this aim, we obtain additional states, $\ket{\tilde{\Psi}_0(D)}$ with spin-adapted \gls{dmrg} at bond dimensions $D\in\{200,400,600,800,1000\}$.
We find an optimized orbital basis, OPT$D$, using each of the $\ket{\tilde{\Psi}_0(D)}$.
In the top pannels of Figure~\ref{fig:fes_D} we show $\Delta E = |E(8000) - E(D)|$, the difference between the energies of $\ket{\tilde{\Psi}_0(8000)}$ and $\ket{\tilde{\Psi}_0(D)}$ as well as the most important amplitude squared in $\ket{\tilde{\Psi}_0(D)}$.
After the optimization, we calculate  $|\braket{J'|\tilde{\Psi}_0(8000)}|^2$ with $\ket{J'}$ in  OPT$D$ and show the resulting values in the bottom pannels of Figure~\ref{fig:fes_D}. There we also display $p_0$(OPT) of Table~\ref{tab:FeS_summary} for reference.
Interestingly, these results clearly show that statevectors obtained at a smaller $D$, while inaccurate in energy, already allow to find an orbital basis which substantially improves the overlap. 

\begin{figure}
    \centering
    \includegraphics[width = 0.9 \textwidth]{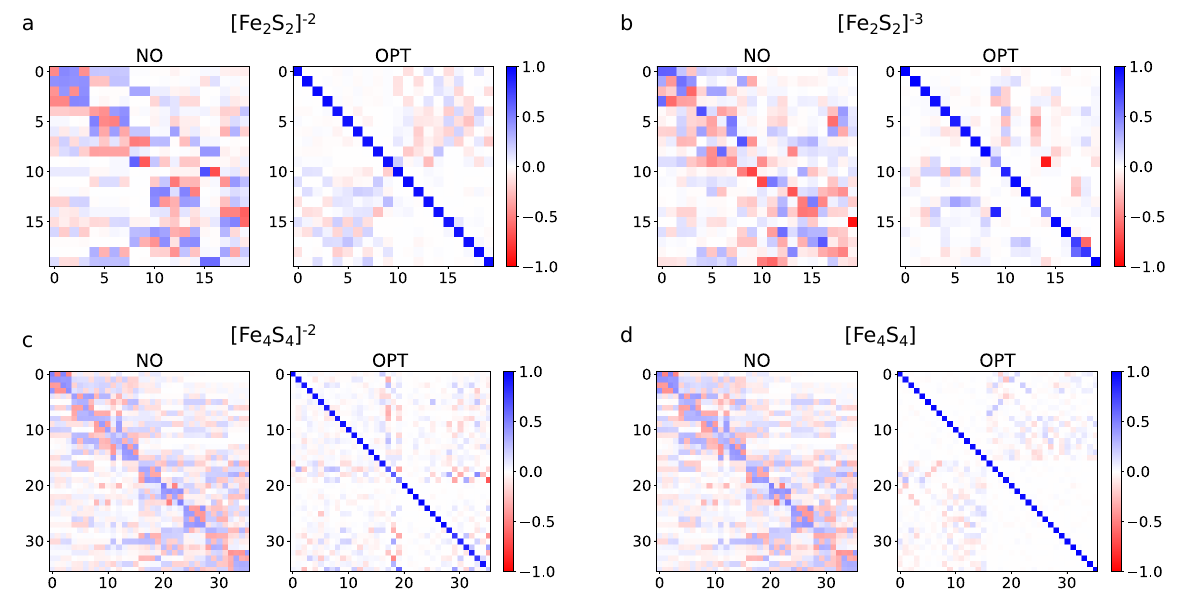}
    \caption{Overlap matrix between the starting MOs and either the NOs or the optimized orbitals (OPT). Plots a, b, c and d are for [Fe$_2$S$_2$]$^{-2}$, [Fe$_2$S$_2$]$^{-3}$, [Fe$_4$S$_4$]$^{-2}$ and [Fe$_4$S$_4$], respectively.}
    \label{fig:fes_orbs}
\end{figure}

Moreover, we compare the \gls{no} and optimized bases. 
The NOs are obtained by diagonalizing the 1-particle density matrix of the $D=8000$ state. 
In Figure~\ref{fig:fes_orbs}, we show the overlap matrix between the starting \glspl{mo} and the \glspl{no} as well as between the \glspl{mo} and the optimized orbitals. As we observed in the acene case, the \glspl{no} and optimized orbitals are different and the latter are much closer to the initial \glspl{mo}. 

Here, the direct calculation of $p_0$(NO) between the Aufbau filled \gls{sd} in the \gls{no} basis, and the \gls{mo} reference state leads to values much lower than $p_0$(MO). 
However, the density of non-diagonal elements in the \gls{mo}-\gls{no} overlap matrix is large. This translates to having more non-vanishing terms in the sum of Eq.~\ref{eq:ovlp_det}. We therefore expect the errors defined in Eqs.~\ref{eq:er_1} and~\ref{eq:er_2} to be large, hence resulting in unreliable $p_0$ estimates. 
To allow for a more fair comparison between the optimized and the \gls{no} basis, we run \gls{dmrg} again in the \gls{no} basis and get $p_0$(NO) as the amplitude squared of the most important determinant in the resulting state. 
This is only done for the two $N_{\text{Fe}}=2$ clusters, as we can afford a large enough bond dimension to compensate for the loss of locality in the orbitals. As the energy and overlap converge at $D=500$, we use this bond dimension to obtain our states in the \gls{no} basis. The results are given in Table~\ref{tab:FeS_summary}.
In both cases we obtain better weights in the optimized basis.

Finally, note that another way of verifying the accuracy of the calculated $p_0$ is to rotate the \gls{mps} underlying $\ket{\tilde{\Psi}_0}$ to the optimized basis and sample from it. However, this procedure was too costly for the systems considered here as $D$ blows up due to the loss of locality in the optimized orbitals. This explains our choice of demonstrating the convergence of $p_0$ with $N_{\text{SD}}$ and $D$ instead.

\section{Cytochrome P450}
\label{sec:P450}
\begin{figure}
    \centering
    \includegraphics[width = 0.5\textwidth]{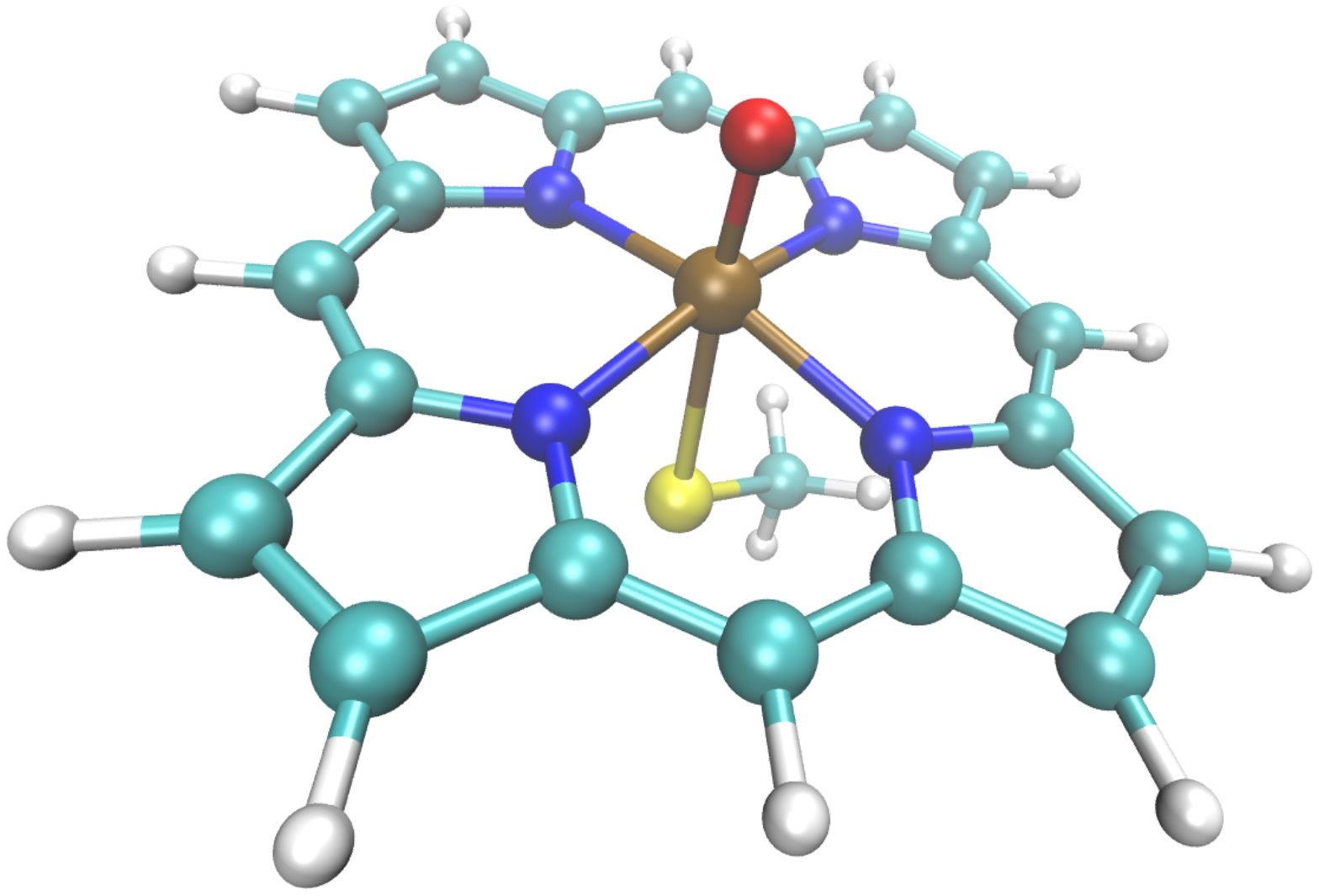}
    \caption{Oxo model of the cytochrome P450 enzyme.}
    \label{fig:oxo_P450}
\end{figure}
The cytochrome P450 enzyme holds significant importance in the pharmaceutical industry.  Recent studies showcased its complex electronic structure and highlight the system as an ideal use case for the first generation of quantum computers~\cite{Goings2022}. The numerical results presented in this paper also underscore that a single determinant already exhibits a considerable overlap with the ground state of P450. However, further enhancing this overlap can reduce the runtime of quantum phase estimation, making the computation more feasible for industrial applications. To evaluate our method's performance on the P450 system, we utilize the O$_2$-bound model (oxo, see Fig.~\ref{fig:oxo_P450}) and active spaces available in~\cite{goings_2022_5941130}. We approximate the ground state with spin-adapted \gls{dmrg} at bond dimension $D=1500$ for the various spins. In all cases we use a singlet-embedding. For both the \quotes{X} (47e,43o) and \quotes{G} (63e,58o) active spaces,  as well as for the $S=1/2$ and $S=5/2$ states, our method yields an approximate 10\% improvement in $p_0$, see Tab.~\ref{tab:p450}. We note that the overlaps reported in Fig.~9 of Ref.~\cite{Goings2022} are considerably higher as the authors report the highest amplitude of a \gls{csf}, while we report the weight (amplitude squared) of a single \gls{sd}. 

\begin{table}[]
    \centering
    \begin{tabular}{|c|c|c|c|c|}
    \hline
        Active space & S & Energy [Ha] & $p_0$(MO) & $p_0$(OPT)\\
        \hline
        G & $5/2$ &  -2756.7356 & 0.53 & 0.58  \\
        G & $1/2$ & -2756.8039 & 0.11 & 0.12 \\
        X & $5/2$ & -2756.9426 & 0.48 & 0.53  \\
        X & $1/2$ &  -2757.0156 & 0.093 & 0.11 \\
        \hline
    \end{tabular}
    \caption{Results for the cytochrome P450 enzyme. The active spaces G and X are taken from Ref.~\cite{Goings2022} and correspond to (47e,43o) and (63e,58o), respectively. The ground state of all systems are approximated with spin-adapted DMRG ($D=1500$).}
    \label{tab:p450}
\end{table}
\end{document}